\documentclass[11pt,twoside]{article}
\usepackage{asp2006}
\usepackage{graphics}
\usepackage{graphicx}
\begin{document}
\title{FLCT: A Fast, Efficient Method for Performing Local Correlation
Tracking}
\author{G.~H. Fisher\altaffilmark{1} and B.~T. Welsch\altaffilmark{1}}

\altaffiltext{1}{Space Sciences Laboratory \#7450, 
7 Gauss Way, University of California, Berkeley, CA 94720-7450}

\begin{abstract}
We describe the computational techniques employed in the recently
updated Fourier local correlation tracking (FLCT) method.  The FLCT
code is then evaluated using a series of simple, 2D, known flow
patterns that test its accuracy and characterize its errors. 
\end{abstract}

\section{Introduction - What is the basic concept behind local correlation
tracking (LCT)?}

Given a pair of 2D maps (``images'') $I_1 (x,y,t)$ and $I_2
(x,y,t2=t1+\delta t)$ of some scalar quantity, with the second image
taken slightly later than the first one, what is the 2-d flow field $(
v_x[x,y], v_y[x,y] )$ which, when applied to the scalar field in the
first image, will most closely resemble the second image?  This
definition of LCT is not a precise one, and the LCT technique
incorporates no physical conservation laws.  Schuck (2005)
\nocite{GF_Schuck2005} showed that LCT methods are more consistent with
an advection equation, rather than a continuity equation.  Here, we
acknowledge these shortcomings and forge ahead, noting that LCT
results must be interpreted carefully.

In Solar Physics, the idea for LCT is generally attributed to
\nocite{GF_November1988} November and Simon (1988).  In the
engineering literature, the problem is known as the ``optical flow''
problem (see Schuck [2006] \nocite{GF_Schuck2006} and references
therein).  Here, we present the technique behind the recently upgraded
Fourier LCT (FLCT) method, first described in Welsch, Fisher, \&
Abbett (2004). \nocite{GF_Welsch2004}

\section{The Mathematical Approach Used by FLCT}

To construct a 2D velocity field that connects two images $I_1 (x,y)$
and $I_2 (x,y)$ taken at two different times $t_1$ and $t_2$, one must
start from some given location within both images, compute a velocity
vector, and then repeat the calculation while varying that location
over all pixel positions.  This involves three high-level operations:
(1) windowing the input images to isolate the neighborhood around the
pixel of interest; (2) computing the correlation function between the
two images; and (3) locating the peak of the cross correlation function.

For each pixel at which a velocity is to be computed, a windowing function
is used to de-emphasize parts of the image far away from that pixel.
FLCT does this localization by multiplying each of the two images by a
Gaussian of width $\sigma$, centered at pixel location $(x_i ,y_j )$.
We denote the resulting images as ``sub-images'' $S_1$ and $S_2$.  The
expressions for $S_1$ and $S_2$ are:
\begin{eqnarray}
S_1^{(i,j)}(x,y) &=& I_1(x, y ) e^{-[(x-x_i )^2+(y-y_j )]^2 / \sigma^2} 
\nonumber \\
S_2^{(i,j)}(x,y) &=& I_2(x, y ) e^{-[(x-x_i )^2+(y-y_j )]^2 / \sigma^2} 
~. \label{eqn:subimages} \end{eqnarray}

The quantity $\sigma$ is a free parameter in FLCT, and its optimal
value changes depending on the nature of the image and the size scales
present in the velocity field. 
As discussed in Welsch {\em et al.}~(2007), \nocite{GF_Welsch2007} one
way to choose the optimal $\sigma$ for a given application is to
select the $\sigma$ that statistically minimizes the difference
between the final image and the advected initial image, $|I_2 - \delta
t (\mathbf{v} \cdot \nabla) I_1|.$

For the $(i,j)$th pixel, the cross-correlation function of 
sub-image 1 with sub-image 2 is defined by
\begin{equation}
C^{i,j}(\delta x, \delta y) = \int \int dx \, dy \,
{S^{i,j}_1}^* (-x, -y) 
S^{i,j}_2 (\delta x-x,\delta y - y) 
~. \label{eqn:crosscor}
\end{equation}

We want to find, for each pair of sub-images $S_1$ and $S_2$ centered
at position $(x_i,y_j)$, the shifts $\delta x$ and $\delta y$ that maximize
$C(\delta x, \delta y)$.  The amplitude of the shifts, divided by the
time $\delta t =t_2-t_1$ between images 1 and 2 defines the velocity
determined by FLCT: $v_x=\delta x/\delta t$, and $v_y=\delta y/\delta t$.

FLCT uses the convolution theorem to compute $C(\delta x,\delta y)$
using Fourier transforms.  If we write ${\cal F}(S_1)= s_1(k_x,k_y)$
and, ${\cal F}(S_2) = s_2(k_x,k_y),$ where ${\cal F}$ denotes Fourier
transform, then the above equation can be written
\begin{equation}
C^{i,j}(\delta x, \delta y) = {\cal F}^{-1}(s_1^* s_2) 
~, \label{eqn:convol}
\end{equation}
where ${\cal F}^{-1}$ denotes the inverse Fourier transform.  We
sometimes find it useful to perform a low-pass Gaussian filter on the
functions $s_1$ and $s_2$ before applying equation (\ref{eqn:convol})
if the original images are noisy.  Other researchers have used
different cross-correlation functions; some calculate the correlation
in $x,y$-space, e.g., November and Simon (1988). \nocite{GF_November1988}

Next, we must find the peak of the cross-correlation function.
Sub-pixel accuracy is required, as shifts are frequently substantially
less than 0.1 pixel.  As a practical matter, we find the peak of
$f(\delta x,\delta y) = |C^{i,j}(\delta x,\delta y)|$, rather than $C(\delta
x,\delta y)$ so that the operation does not involve complex
arithmetic.  For notational simplicity, we henceforth use $x,y$ for
$\delta x,\delta y$ in the following discussion.

Previous versions of FLCT followed \cite{GF_November1988}, and
interpolated $f$ around its pixel-accuracy peak onto a finer grid, and
took the location of the maximum of the resulting discretely sampled
function as the shift.  This approach, however, is only as accurate as
the resolution of the interpolated grid, and is therefore
computationally expensive --- many unnecessary interpolations are
required to reach the necessary spatial resolution.

Our current version employs a curve-fitting approach to find the peak
in $f(x,y)$ that was inspired by that of Chae (2004, private
communication; LCT code written in IDL).  First, since the images and
sub-images are computed at discrete points in space, we identify the
pixel coordinates $(x_m,y_n)$ of the largest value of $f$, denoting
the largest value of $f$ as $f(x_m,y_n).$ Note that $(x_m,y_n)$ may
not be equal to $(x_i,y_j)$, if the location of the peak of $f(x,y)$
has shifted by more than a pixel in $x$ or $y.$ To find the peak to
sub-pixel resolution, we then Taylor-expand $f(x,y)$ to 2nd order
about the $(x_m,y_n)$ location, denoting the expansion as $f_T(x,y)$,
\begin{eqnarray}
f_T(x,y) &\equiv& f(x_m, y_n) 
+ \frac{\partial f}{\partial x}(x - x_m)
+ \frac{\partial f}{\partial y}(y - y_n) \\ \nonumber
&+& \frac{\partial^2 f}{\partial x \partial y}(x - x_m)(y - y_n)
+ \frac{1}{2}\frac{\partial^2 f}{\partial x^2}(x - x_m)^2
+ \frac{1}{2}\frac{\partial^2 f}{\partial y^2}(y - y_n)^2
~, \label{eqn:taylor} \end{eqnarray}
where the partial derivatives are evaluated at the point $(x_m, y_n)$.

At the peak, we require that the $x$ and $y$ partial derivatives of the
Taylor expansion $f_T(x,y)$ vanish. These conditions result in a
pair of linear equations which allow us to solve for the location
$(x_{\rm max},y_{\rm max})$ of the peak:
\begin{eqnarray}
x_{\rm max} - x_m &=& 
\left ( \frac{\partial^2 f}{\partial y^2} 
\frac{\partial f}{\partial x}
-
\frac{\partial^2 f}{\partial x \partial y}
\frac{\partial f}{\partial y}
\right ) 
\left ( \left ( 
\frac{\partial^2 f}{\partial x \partial y}
\right)^2
-
\frac{\partial^2 f}{\partial x^2}
\frac{\partial^2 f}{\partial y^2}
\right )^{-1} \label{eqn:maxx} \\
y_{\rm max} - y_n &=& 
\left ( \frac{\partial^2 f}{\partial x^2} 
\frac{\partial f}{\partial y}
-
\frac{\partial^2 f}{\partial x \partial y}
\frac{\partial f}{\partial x}
\right ) 
\left ( \left ( 
\frac{\partial^2 f}{\partial x \partial y}
\right)^2
-
\frac{\partial^2 f}{\partial x^2}
\frac{\partial^2 f}{\partial y^2}
\right )^{-1} 
~. \label{eqn:maxy} \end{eqnarray}
To evaluate the partial derivatives, we use standard, second-order 
finite difference expressions, assuming a uniform grid in the 
sub-images:
\begin{eqnarray}
\frac{\partial f}{\partial x} = (f_{m+1,n} - f_{m-1,n})/2 \Delta x &,& 
\frac{\partial f}{\partial y} = (f_{m,n+1} - f_{m,n-1})/2 \Delta y \nonumber \\
\frac{\partial^2 f}{\partial x^2} = (f_{m+1,n} + f_{m-1,n} - 2 f_{m,n})/
\Delta x^2 
&,&
\frac{\partial^2 f}{\partial y^2} = (f_{m,n+1} + f_{m,n-1} - 2 f_{m,n})/
\Delta y^2 \nonumber \\
\frac{\partial^2 f}{\partial x \partial y} = 
(f_{m+1,n+1} + f_{m-1,n-1} &-& f_{m-1,n+1} - f_{m+1,n-1})/
4 \Delta x \Delta y 
~. \label{eqn:derivs} \end{eqnarray}
The total shift in $(x,y)$ that corresponds to the peak in $f$ 
is then given by 
\begin{equation}
\delta x = (x_m - x_i) + (x_{\rm max} - x_m) 
~,~ 
\delta y = (y_n - y_j) + (y_{\rm max} - y_n)
~. \label{eqn:shifts} \end{equation}
Equations (\ref{eqn:maxx}) through (\ref{eqn:shifts}) thus provide a 
simple and accurate method for finding the peak of the cross-
correlation function to sub-pixel resolution.

\section{The Computational Approach Used by FLCT}
 
Our method requires two input images, $I_1$ and $I_2$, recorded at
times $t_1$ and $t_2=t_1+dt$.  The algorithm then loops over all pixels
in $I_1$ and $I_2$ for which $|I_1 + I_2|/2 > I_{\rm thr}$, where
$I_{\rm thr}$ is a user-set threshold.  For each such pixel, it must:
\begin{enumerate}
\item Compute sub-images $S_1$ and $S_2.$ 
\item Compute Fourier transforms of $S_1$ and $S_2$, $s_1$ and $s_2.$ 
\item Perform low-pass filters on $s_1$, $s_2$ if needed. 
\item Compute inverse Fourier transform of $s_1^* s_2$.  This is $C^{i,j}$.
\item Compute the absolute value of $C^{i,j}, |C^{i,j}|$.
\item Compute the shifts $\delta x, \delta y$ that maximize $|C^{i,j}|$.
\item Compute velocities $v_x=\delta x/\delta t, v_y=\delta y/\delta t.$
\end{enumerate}

The FLCT code, which is currently called \texttt{flct}, was initially
written in IDL, but has since been re-written in C for portability
and speed.  FLCT also uses its own endian-independent binary input
format for the images and for the output velocity arrays.  FLCT uses
the FFTW library (version 3) for computing the Fast-Fourier Transforms
needed to compute the cross-correlation functions.  We have written
IDL procedures which write the two images \texttt{flct} needs into an input
file, and which read the flct output file, consisting of
two-dimensional, floating point arrays of $v_x,v_y$, and a mask array,
$v_m,$ equal to one in pixels with $|I_1 + I_2|/2 > I_{\rm thr}$ and
zero elsewhere.  These IDL i/o utilities enable \texttt{flct} to be run
easily from within an IDL session.

\section{Running FLCT}

To compile the FLCT code, the only external library needed is the
FFTW3 library.  To download a copy of the FLCT source code and
compilation instructions, along with IDL i/o procedures, go to our web
site.\footnote[1]{\texttt{http://solarmuri.ssl.berkeley.edu/overview/publicdownloads/software.html}}
Be sure to get the C version (not the IDL version), and get version
\texttt{test\_13} or later.

The \texttt{flct} is code invoked from the command-line -- either a
terminal window on a Unix-like machine, from the command-prompt tool
in MS Windows, or from within an IDL session using either a shell
escape character or the spawn command.  In all cases, the syntax for
flct is as follows:

\texttt{flct infile outfile deltat deltas sigma -t thr -k kr -h -q} \\

\noindent
{\bf Required Arguments:} 

\begin{list}{}{}
\item \texttt{infile} - Contains 2 images for local correlation tracking.
To create \texttt{infile} use the IDL procedure
\texttt{vcimage2out.pro}

\item
\texttt{outfile} - contains output $v_x, v_y, v_m$ ($x,y$ velocity and
mask arrays).  Mask array is zero where velocity not computed. To read
\texttt{outfile}, use the IDL procedure \texttt{vcimage3in.pro} 

\item
\texttt{deltat} - Amount of time between images.

\item
\texttt{deltas} - Units of length of the side of a single pixel; velocity
is computed in units of \texttt{deltas/deltat}.

\item
\texttt{sigma} - Sub-images are weighted by Gaussian of width sigma.  If
sigma is set to 0, only single values of shifts are returned as $v_x,
v_y.$  These values correspond to the overall shifts between the two
images, and these values can be used to co-register the images if
desired.
\end{list}

\noindent
{\bf Optional parameters:}
\begin{list}{}{}
\item \texttt{-t thr} - Determines threshold parameter $I_{\rm thr}$.  If
$|I_1 + I_2|/2 < I_{\rm thr}$ in a pixel, then skip calculation of
shifts for that pixel.  For \texttt{thr} $\in (0,1),$ \texttt{thr} is
assumed to be in relative units of the maximum absolute pixel value in
$\{|I_1|,|I_2|\}$.  To force \texttt{thr} values between zero and unity
to be in absolute units, append an 'a' to \texttt{thr}.  If
calculation is skipped for a pixel, then the $v_m$ mask array is
set to zero at that pixel location.  It is otherwise set to 1.
\item \texttt{-k kr} - Apply a low-pass filter to the sub-images, with a
 Gaussian of a characteristic wavenumber that is a factor of $k_r$
 times the largest possible wave numbers in $x,y$ directions.  $k_r$
 should be positive.  (If $kr >> 1$, the unfiltered result should be
 recovered). This option is sometimes useful for datasets that contain
 noise at high spatial frequencies.  Test cases in which MDI Quiet-Sun
 magnetograms are shifted by 10 milli-pixels frequently result in a
 factor of two under-estimate by \texttt{flct} of the applied shift.  By
 setting $k_r$ to numbers between 0.2 to 0.5, the full applied shifts
 can be recovered.  Experimentation is recommended.
\item \texttt{-h} - ``hires'' mode. Set this flag to use cubic convolution
 interpolation of the $f(x,y)$ function to re-grid it to higher
 resolution before finding shift.  It is unclear that this
 interpolation improves accuracy. Currently, this option seldom used, and 
 may be dropped in the future.  It has been retained for heritage.
\item \texttt{-q} - Flag to suppress printing of all non-error messages.
\end{list}

From within IDL, a simple demonstration of FLCT can be run using the
following commands. \\
\\
\texttt{IDL$>$ f1=randomu(seed,101,101) } \\
\texttt{IDL$>$ f2=shift(f1,1,-1) } \\
\texttt{IDL$>$ vcimage2out,f1,f2,`testin.dat' } \\
\texttt{IDL$>$ \$flct testin.dat testout.dat 1. 1. 15. } \\
\texttt{IDL$>$ vcimage3in,vx,vy,vm,`testout.dat' } \\
\texttt{IDL$>$ shade\_surf,vx } \\
\texttt{IDL$>$ shade\_surf,vy } \\

\section{FLCT Speed}

How quickly does FLCT run?  On a 1.6 GHz Pentium M (laptop processor)
running MS Windows XP, the FLCT code flct can process roughly 200
pixels per second for images with over 40,000 pixels (time per pixel
is much less than this for smaller images).  The code is speeded up
dramatically if the \texttt{-t thr} option is used, in which the
velocity calculation is skipped if the image values are less than the
chosen threshold.  There are many avenues available for further
speedup of the FLCT code.  Very little experimentation has been done
with multiple threads thus far.  An experimental version of flct
written in Fortran, using MPI has been tested on our Linux cluster,
\textsl{grizzly}, and the compute speed was found to scale almost
perfectly with the number of processors.

\begin{figure}[!ht]
\includegraphics[width=5.5in]{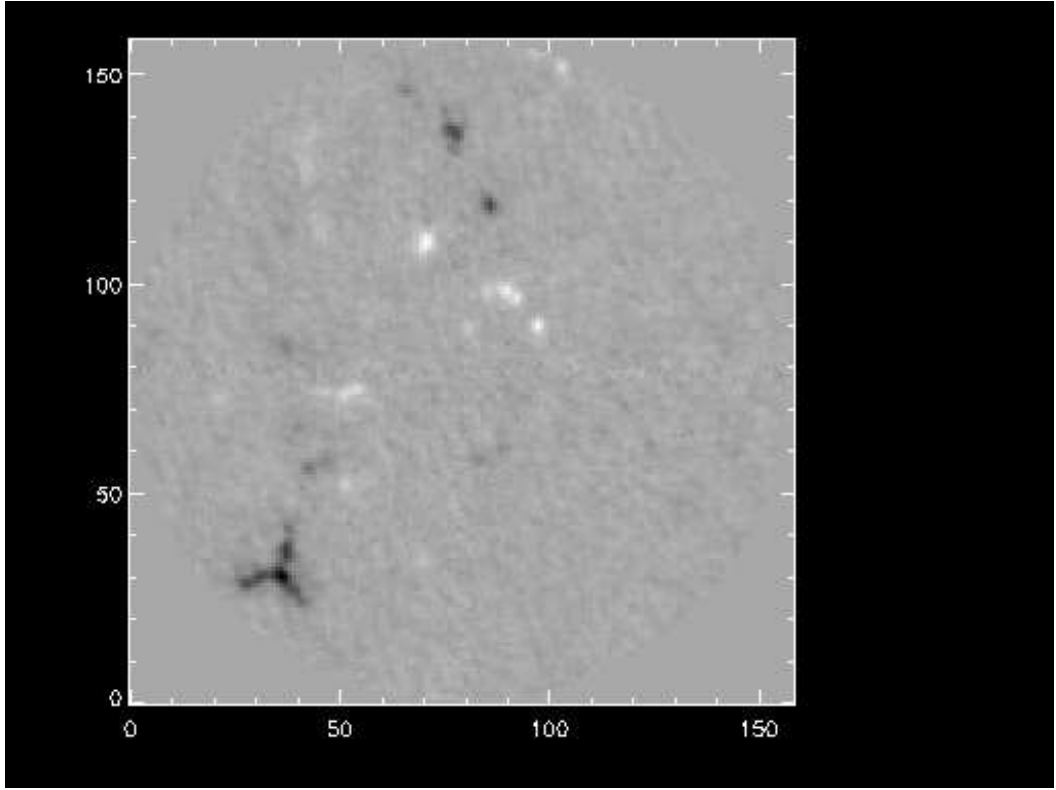}
\caption{\label{fig:mdiqs} An MDI magnetogram of the Quiet Sun.  By
applying known shifts and rotations to such magnetograms, we can test
the performance of FLCT.}
\end{figure}

\section{Tests of FLCT}

Figure \ref{fig:mdiqs} shows an MDI \cite{GF_Scherrer1995} magnetogram of
the Quiet Sun.  (For a discussion about applying tracking techniques
like FLCT to magnetograms, see the paper Welsch and Fisher in this
volume.)  To test the ability of FLCT to reconstruct rotations, we
rotated the image, using cubic-convolution interpolation, by a fixed
angle about the center of the image to generate a second image.  The
applied ``velocity field'' then corresponds to simple, solid-body
rotation, with the speed increasing linearly with radius, $r$, from
image center.

\begin{figure}[!ht]
\includegraphics[width=5.5in]{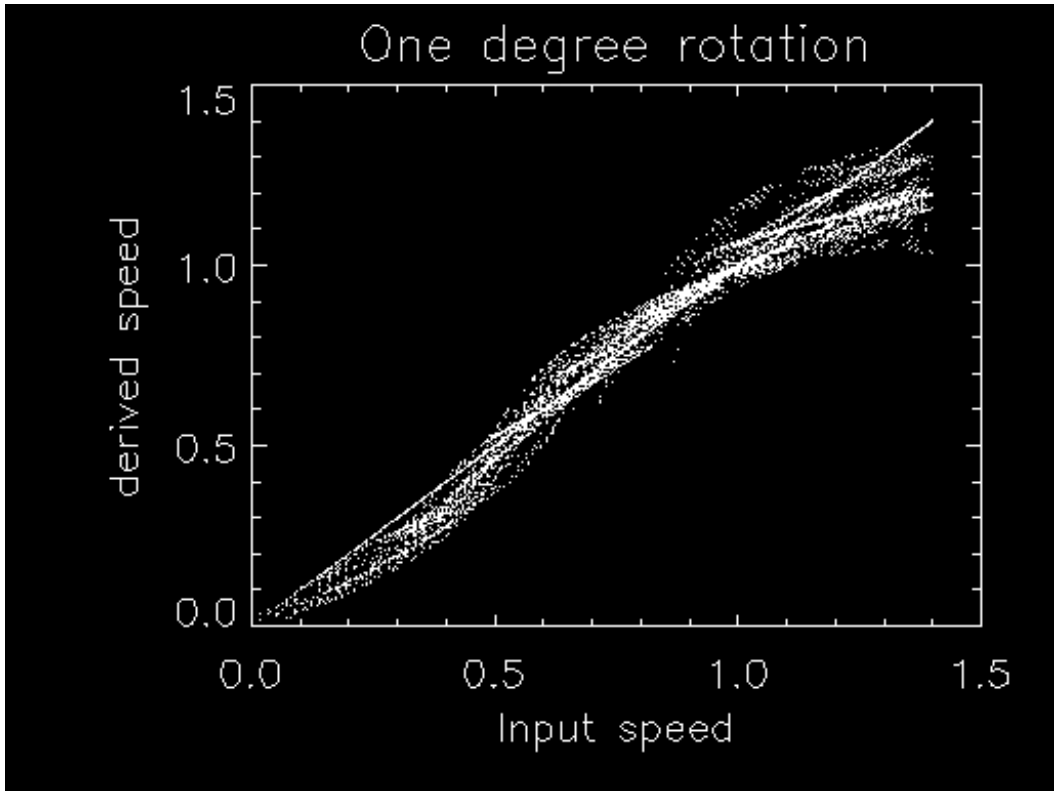}
\caption{\label{fig:1deg} A scatter plot of derived speeds versus
input speeds, for an applied rotation of 1$^\circ$. The straight line
shows the applied speed. There is no significant bias in the derived
speed.}
\end{figure}

\begin{figure}[!ht]
\includegraphics[width=5.5in]{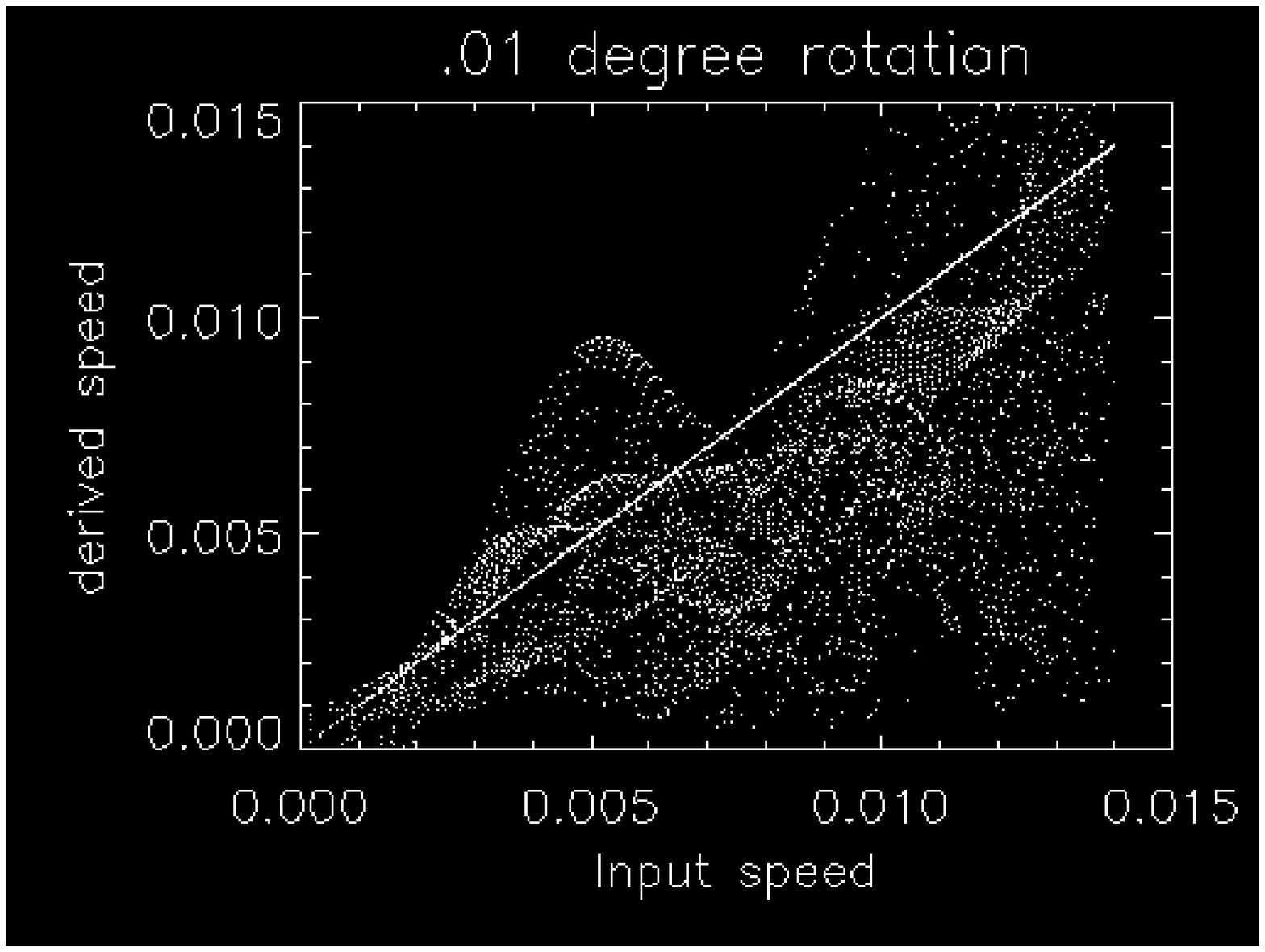}
\caption{\label{fig:hundeg} A scatter plot of derived speeds versus
input speeds, for an applied rotation of 0.01$^\circ$. The scatter is
significant, and the derived speeds are biased on the low side as
compared to the applied speeds.}
\end{figure}

How well does FLCT recover the applied velocity field?  In Figures
\ref{fig:1deg} and \ref{fig:hundeg}, we show scatter plots of derived
speeds versus input speeds for a rotations of 1$^\circ$ and
0.01$^\circ$, respectively.  Units of velocity are in pixel widths per
unit time.  The straight lines show the applied speed, and the scatter
plots show the inverted speeds for all inverted pixels (only points
with $|B| >$ 10 G were inverted).  $\sigma =$ 15 pixels was assumed.
For the 1$^\circ$ rotation, the scatter is typically 0.1-0.2, and
there is no significant bias in the derived speed.  For the
0.01$^\circ$ rotation, the scatter is significant, and the derived
speeds are biased on the low side as compared to the applied speeds.
For the 1$^\circ$ rotation, the recovered rotation profile (not shown)
accurately reproduces the imposed rotation profile.  These and other
tests suggest that FLCT can accurately recover shifts $> 0.1$ pixels,
but smaller shifts can be problematic.

\section{Discussion}

What is the difference between the new (upgraded) FLCT code, described
here, and that originally described in Welsch {\em et al.}~(2004)?
\nocite{GF_Welsch2004} The main difference is in how the location of the
peak of the cross-correlation function is determined to sub-pixel
resolution.  The old version used cubic convolution interpolation to
compute the cross-correlation function on a much finer grid (0.1 or
0.02 pixel spacing) and then simply found the location of the largest
value within the finer grid.  We (and other users) found that this
resulted in serious ``quantization'' errors when the time between
images was small enough that shifts were of order 0.1-0.2 pixels or
less.  The new code uses equations (\ref{eqn:derivs}) and
(\ref{eqn:shifts}) of this paper to find the location of the peak to
sub-pixel resolution.  We have found this to be more accurate, and
more computationally efficient --- and therefore much faster --- than
the old technique.  We strongly recommend that users use the new
version and not any of the old versions.

\acknowledgements{This work was supported by NSF grant ATM-064130,
NASA Heliophysics Division through funding from the SR\&T and Theory
Programs, by NSF-ATM through the SHINE program, and by the DoD MURI
grant, "Understanding Magnetic Eruptions on the Sun and their
Interplanetary Consequences".  }



\end{document}